\documentclass[aps,prl,twocolumn,showpacs,superscriptaddress]{revtex4}
\usepackage{natbib}
\usepackage{epsfig,xcolor,graphics}
\usepackage{amsmath, mathrsfs}
\usepackage{amssymb}
\usepackage{siunitx}
\begin{document}
\title{Cutting and Slicing Weak Solids}
\author{Serge Mora}
\affiliation{Laboratoire de M\'ecanique et G\'enie Civil, Universit\'e de Montpellier and CNRS, France}
\email[Corresponding author: ]{serge.mora@umontpellier.fr}
\author{Yves Pomeau}
\affiliation{LadHyX, UMR CNRS 7646, Ecole Polytechnique, 91128 Palaiseau, France}
\email{yves.pomeau@gmail.com}
\date{\today}
\begin{abstract}
  Dicing soft solids with a sharp knife is quicker and smoother if the blade is sliding rapidly parallel to its edge in addition to the normal squeezing motion. We explain this common observation with a consistent theory suited for soft gels and departing from the standard theories of elastic fracture mechanics developed for a century. The gel is assumed to locally fail when submitted to stresses exceeding a threshold $\sigma_1$. The changes in its structure generate a liquid layer coating the blade and transmitting the stress through viscous forces. The driving parameters are the ratio $U/W$ of the normal to the tangential velocity of the blade, and  the characteristic length  $\eta W/\sigma_1$, with $\eta$ the viscosity of the liquid. The existence of a maximal value of $U/W$ for a steady regime explains the crucial role of the tangential velocity for slicing biological and other soft materials. 
\end{abstract}
\maketitle

\label{section:order}
Cutting soft materials has been done forever and is nowadays encountered in a large range of processes, from precision surgery \cite{Sturm2008} to histology \cite{Slaoui2010} to food industry \cite{Kamyab1998,Atkins2004}. It also occurs in everyday life, where it is commonly observed that cutting cheese or meat is made much easier by sliding rapidly the knife \cite{Atkins2006}. Pulling a sharp knife on an elastic material yields a very large stress at the tip because a finite force (per unit length along the edge) is exerted at the contact line. A supplementary sliding of the knife along this contact line produces an additional shear stress. The resulting tensile stress is far more efficient than the compressive stress produced by the single normal displacement of the knife. Hence, cutting soft materials is facilitated by the sliding, requiring a much lower normal force compared to a simple compression \cite{Reyssat2012}.
Beyond this base explanation, the origin of a critical tensile stress needed for cutting a soft gel, as well as the effect of the tangential velocity of the blade, must be elucidated. The explanation to this intriguing question necessarily comes from the specific properties shared by the class of material considered here, {\it i.e.} soft gels. It must be consistent with the large deformations these materials can usually withstand before being cut. In addition, due to the low values of the stresses involved during the failure of these weak solids, macroscopic length-scales emerge, for instance for the energy dissipation zone \cite{Foyart2016,Lefranc2016}. Hence, a consistent description of the dissipative zone is required to properly describe the mechanics of the cutting. Indeed, the effect of the velocity of the blade is likely to be closely related to the rate of energy dissipation. For these reasons, the physics of slicing soft gels with a sharp wedge has little chance of being well captured by the overspread theory of the linear elastic fracture mechanics, or even its extension to the large deformations \cite{Hui2015,Tabuteau2011}. 

Here, we give a coherent explanation of the role of the tangential velocity of the blade during the slicing of a soft gel based on a self-consistent theory. The basis of the idea is as follows. We think to a model material that can withstand stress until a given maximum value, called $\sigma_1$ here and defined more accurately later. For stresses lower than this maximum, the material under consideration can return reversibly to its unperturbed state when unstressed. For stresses larger than $\sigma_1$, the material melts like a solid above the melting temperature. This is a reasonable assumption for gels because gels are made of a network filling a liquid (usually water). Once a big enough proportion of links is broken, the network disappears and the gel becomes a liquid suspension of small particles, without the cohesiveness of a solid.
Obviously, this does not take into account a likely transition from solid to liquid through a kind of intermediate state, which certainly takes place, but we assume that the thickness of the transition region is small enough to be neglected, at least in a first approximation. Moreover, to simplify the matter, we shall consider steady situations only.
The material can melt reversibly or irreversibly. If it does it reversibly, it heals once the stress decreases below $\sigma_1$, recovering its elastic properties \cite{Ligoure2013,Coussot2014}. If not, it remains permanently melted whatever the forthcoming changes of the stress \cite{Coussot2014}. It is also possible that the gel cannot stand stress beyond a critical value without expelling some of its liquid part by permeation \cite{Scherer1998}. Most likely the two phenomena (melting and permeation) occur simultaneously in a stressed gel. The different possible mechanisms inducing the formation of the liquid layer do not make a difference with the cutting mechanism considered here, and so we consider these possibilities as a single one.

\begin{figure}[!h]
\begin{center}
  \includegraphics[width=0.45\textwidth]{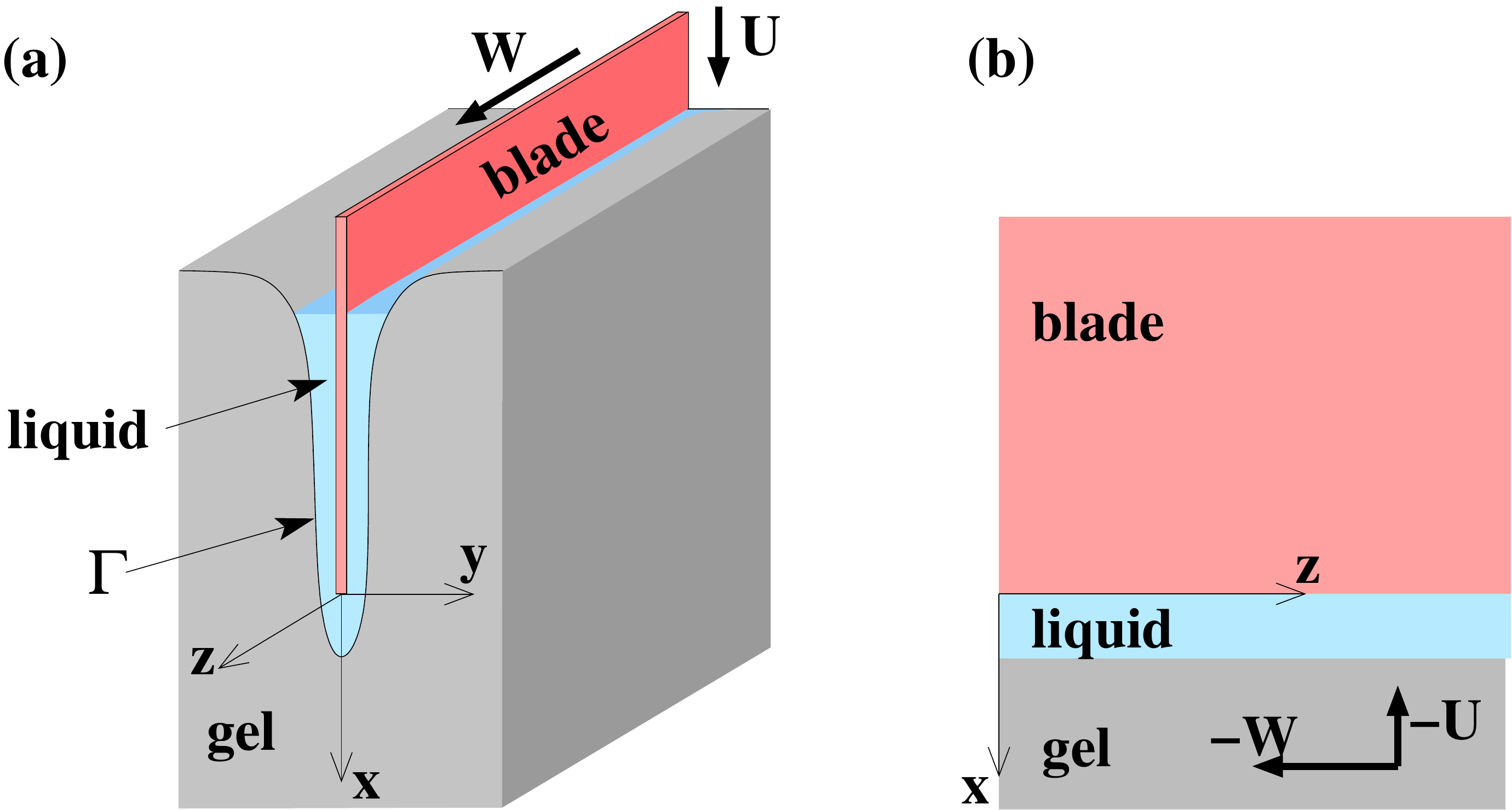}
\end{center}
\caption{(a) Sketch of a blade moving at velocity $U$ in the cutting (normal) direction $x$, and velocity $W$ in the sliding direction $z$ (parallel to the initial surface of the gel). The shape of the liquid-gel interface in plane ($yz$) is represented by the curve $\Gamma$. (b) View in plane ($xz$): in the frame moving with the blade, gel velocity is $-U$ along $x$, and $-W$ along $z$. The system is assumed to be invariant along $z$.} \label{fig : scheme 3d}
\end{figure}

The physical explanation of why it is easier to cut soft materials by sliding the knife rapidly parallel to itself follows from our assumption that the material to be cut cannot stand stresses beyond a critical value. Such a stress is transmitted from the knife to the gel by a thin viscous layer of liquid made either by the broken gel and/or by expelled liquid (Fig.~\ref{fig : scheme 3d}). This stress depends on viscosity, knife velocity, and the geometry of the layer. It includes a component coming from the forward cutting motion of the knife toward it, but also a component due to the sliding motion, usually at a far bigger velocity than in the cutting direction.\\ 

\begin{figure}[!h]
\begin{center}
\includegraphics[width=0.4\textwidth]{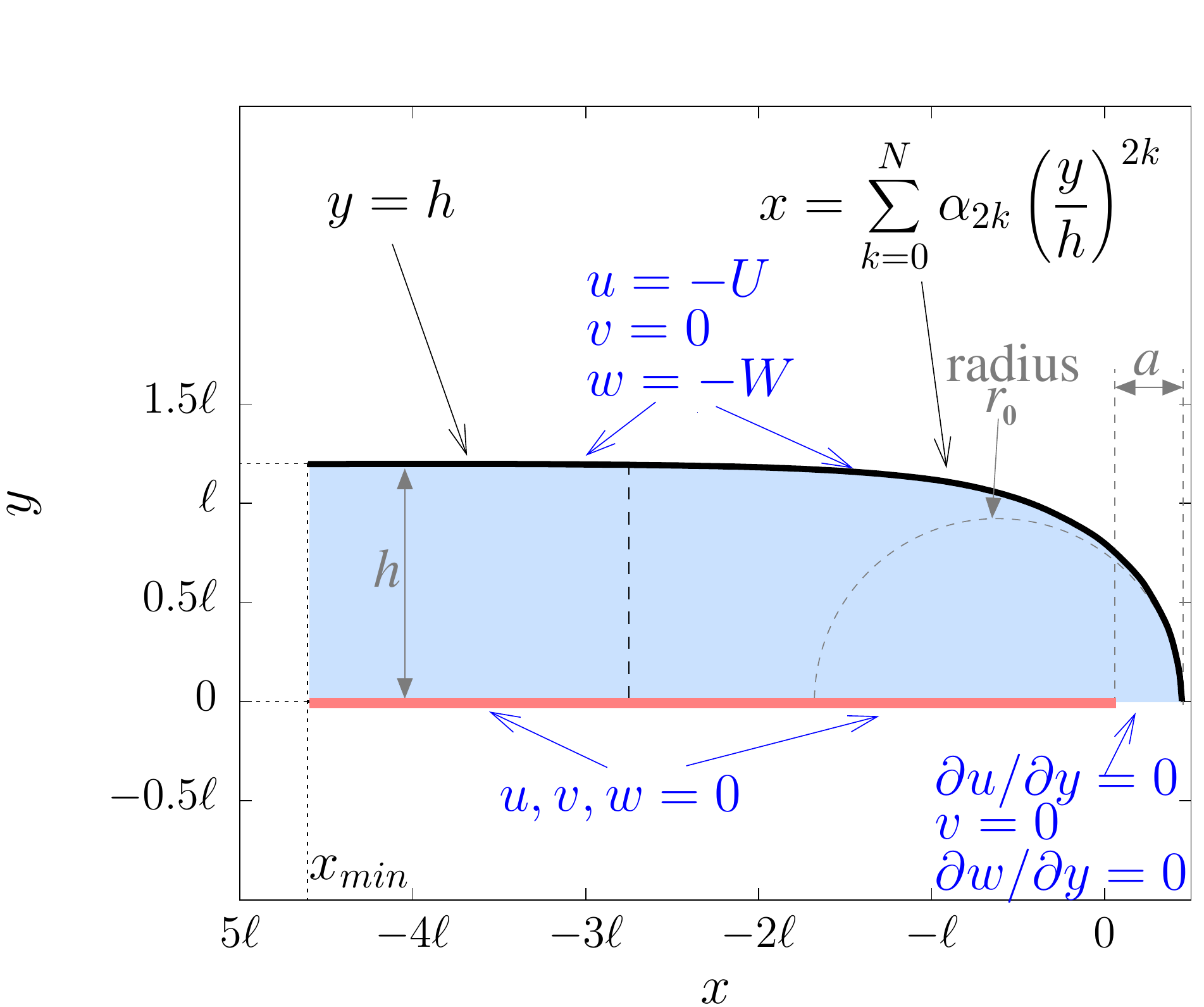}
\end{center}
\caption{Domain $\Omega$ defined by Eq.~\ref{eqn : Gamma} for the computation of the velocities (colored in light blue). Black solid line is the liquid-gel interface, referred as $\Gamma$. The blade is in pink. $a$ is the distance between the edge of the blade and the tip of the notch. $r_0$ is the radius of curvature at the tip, and $h$ is the thickness of the liquid layer far from the tip. The boundary conditions (marked in blue) are $u,v,w=0$ for $y=0$ and  $x_{min}<x<0$; 
 $\partial (u,w)/\partial y=0$ and $v=0$ for $y=0$  and  $0<x<\Gamma(0)$; $u=-U$, $v=0$ and $w=-W$ for  $x=\Gamma(y)$ or $y=h$.} \label{fig : mesh}
\end{figure}

We consider a thin blade consisting of a sharp wedge with an infinitely small angle so that its geometry can be approached, in a first approximation,  as a half plane of zero thickness. This half plane (the blade) is assumed to move parallel to its own plane both in the $z$ direction (the sliding motion) and inward a gel in the $x$ direction perpendicular to the edge (Fig.~\ref{fig : scheme 3d}). Lastly, $y$ is the direction perpendicular to the plane of the knife. Let $U$ be the velocity of the knife in the direction of cutting and $W$ the velocity of slicing in the $z$ direction. We consider a steady state, namely when the knife has advanced for a sufficiently long time to have made a self-reproducing groove in the gel ahead of it. Therefore, the half plane $y = 0$, $x < 0$ made by the idealized knife advances at zero speed in the frame moving with the knife, whereas the gel advances at speed $(-W)$ in the sliding direction and $(-U)$ toward the knife. The half plane is coated by the layer of viscous liquid created by the destruction of the gel and/or by permeation of its liquid component. In this liquid the fluid velocity is assumed to be small enough to make valid the linear Stokes equation. The fluid velocity can be split into two components: the velocity component $(u,v)$ in the $(x, y)$ plane and the velocity component in the $z$ direction, $w$ \cite{Landau1984}.  From the incompressibility condition,
\begin{equation}
  u_{,x}+v_{,y}=0,
  \label{eqn : incompressibility}
\end{equation}
where $u_{,x} =  \frac{ \partial u}{\partial x }$,  etc., Stokes equation is
\begin{eqnarray}
  \eta (u_{,xx}+u_{,yy})&=&p_{,x}, \label{eqn : stokes u}\\
  \eta (v_{,xx}+v_{,yy})&=&p_{,y}, \label{eqn : stokes v}\\
  w_{,xx}+w_{,yy} &=& 0, \label{eqn : stokes w}
\end{eqnarray}
with $p$ the pressure and $\eta$ the shear viscosity of the liquid. Now we have to give the boundary conditions. One set of boundary conditions is imposed on the surface of the knife, namely on the half plane $y = 0$, $x< 0$. In the frame of reference of the knife, the fluid velocity is just zero on this boundary (see Fig.~\ref{fig : scheme 3d}). This amounts to imposing $(u,v,w) = (0,0,0)$ on $ y = 0$, $x< 0$. Let $ \Gamma$ be the curve making the cross section of the interface between the gel and the liquid layer. This is a line in the plane $(x,y)$, and we impose the continuity of the flux of matter across this line and the continuity of stress. Let us consider first the continuity of matter. This is done by imposing that when the gel transforms into liquid, mass is conserved: the flux of matter on the gel side is equal to the flux of liquid on the other side. Because the mass density is almost the same on both sides of $ \Gamma$, continuity of the flux of matter is equivalent to the continuity of velocity. On the side of the gel, the velocity is just the velocity of the frame of reference of the knife. In Cartesian coordinates this is the velocity $(-U, 0, -W)$. This also makes the boundary conditions for the velocity field in the liquid on $ \Gamma$: $ u \vert_{\Gamma} = - U$, $ v \vert_{\Gamma} =  0$, and $w \vert_{\Gamma} =  -W$ (Fig.~\ref{fig : mesh}). Given $\Gamma$, this yields the right number of conditions for the problem.

There remains to get the condition that determines $\Gamma$.  This is the condition that the stress there is exactly at the given critical value $ \sigma_1$. Since the gel is mainly composed of solvent, the same as the liquid phase, the interfacial tension between the two phases is negligible in a first approximation \cite{Mora2013}, and the stress has to take the same value on the gel side and on the liquid side of the boundary. The Cauchy stress tensor in the liquid phase is $\mathbf{\underline{\underline{{\sigma}}}}=\mathbf{\underline{\underline{\tilde{\sigma}}}}+ p \mathbb{I}$ with  $\mathbb{I}$ the identity tensor and $\mathbf{\underline{\underline{\tilde{\sigma}}}}$ the deviatoric stress given by the standard formulas for viscous fluids \cite{Landau1984},
\begin{equation}
  \mathbf{\underline{\underline{\tilde {\sigma}}}}=\eta \left(\begin{array}{ccc}
    2u_{,x} & u_{,y}+v_{,x} & w_{,x}\\
    u_{,y}+v_{,x} & 2v_{,y} & w_{,y}\\
    w_{,x} & w_{,y} & 0
    \end{array} \right).  \label{eqn : cauchy tensor}
\end{equation}
There remains to give the expression of the critical stress, $\Sigma(\mathbf{\underline{\underline{\sigma}}})$, that is to be made equal to the critical value $\sigma_1$. For a isotropic solid, this stress has to be independent of the choice of coordinates, and hence to be a function of the invariants of $\mathbf{\underline{\underline{\sigma}}}$. Several expressions are possible, and we consider here the widespread von Mises criterion,  which amounts to assuming that failure occurs as the elastic energy density of distortion reaches a critical value \cite{Hencky1924}. Other criteria, such as the Tresca criterion based on the maximum shear stress or the maximum principal stress criterion \cite{Christensen2016}, 
could be considered as well.
The von Mises stress $\Sigma$, taken here as measure of the stress, is defined by $\Sigma^2 = \frac{3}{2}\mathrm{tr}(\mathbf{\underline{\underline{\tilde{\sigma}}}}^2)$ \cite{Mises1913,Christensen2016}. In the geometry in consideration, this yields the following: 
\begin{equation}
\Sigma^2=3\eta^2 \left(w_{,x}^2+ w_{, y}^2+v_{,x}^2+u_{,y}^2+4u_{,x}^2+2v_{, x}u_{, y}\right).
  \label{eqn : von mises}
  \end{equation}
The equation of the curve $\Gamma$ is found by imposing
\begin{equation}
  \vert \Sigma  \vert = \sigma_1
  \label{eqn : critere}
\end{equation}
on $\Gamma$,  $\sigma_1$  being a given positive quantity with the physical dimension of a stress. 

Taking $\ell=\sqrt{3}\eta W/\sigma_1$ as unit length and $W$ as unit speed, Eqs.~\ref{eqn : incompressibility}-\ref{eqn : stokes w} and Eq. \ref{eqn : critere} can be reformulated with a unique parameter, the ratio $\zeta=U/W$ of the sliding velocity to the cutting velocity. Taking $\sigma_1$ of the order of magnitude of the shear modulus of a soft hydrogel (a reasonable assumption for polymeric gels) {\it e.g.},  $\sigma_1=\SI{100}{\pascal}$, and  $\eta=\SI{10}{\milli \pascal \second}$  (the liquid layer being a mixture of water and molecules resulting from gel breakage), we find, with $W=\SI{1}{\meter \per \second}$ for the tangential velocity, $\ell=\SI{0.1}{\milli \meter}$, a small but measurable macroscopic length. 

In the following, the thickness $h$ of the liquid layer along the blade (far behind the edge), the distance $a$ between the edge of the blade and the tip of the notch, and the radius of curvature $r_0$ of the gel surface at the tip (see Fig.~\ref{fig : mesh}), are computed as functions of $\zeta$ and $\ell$.\\

Let us first consider the special situation in which the blade is pushed normally toward the gel without sliding, $W=0$. Due to mirror symmetry with respect to plane $y=0$, $u_{,y}=0$ and $v=0$ along axis $y=0$ for $x>0$ (see Fig.~\ref{fig : mesh}); hence, Eq.~\ref{eqn : critere} simplifies in $\Sigma^2=12\eta^2 u_{,x}^2$ or equivalently [from Eq.~\ref{eqn : incompressibility}] $\Sigma^2=12\eta^2 v_{,y}^2$ ($x>0$ and $y=0$). The tangent of $\Gamma$ at $(x,y)=(a,0)$ being parallel to $y$ by symmetry, one deduces from $v_{\vert \Gamma} =0$ that $v_{,y}=0$ at $(x,y)=(a,0)$ and, then, that the stress $\Sigma$ at $(x,y)=(a,0)$ has to be equal to zero. Since $\sigma_1 \ne 0$, the condition $\Sigma=\sigma_1$ on $\Gamma$ cannot be fulfilled. This means that no stationary state exists for $W=0$, and highlights the overriding role played by the sliding of the blade.\\
Note that in this case ($W=0$), $\Sigma$ is the unique nonzero scalar invariant of $\underline{\underline{\mathbf{\tilde{\sigma}}}}$ and taking the von Mises criteria among the others criteria does not limit the generality. \\ 

In what follows, the tangential velocity is non zero.      
Far behind the edge of the blade ($x \ll -\ell$), derivatives along $x$ direction are zero and $u(y)$ is parabolic. According to the boundary conditions at $y=0$ and $y=h$ and the incompressibility condition, (the average value of $u$ along $y$ is equal to $-U$), $u=-U\left(4y/h-3y^2/h^2\right)$. This is a simple combination of Couette and Poiseuille flows, the pressure gradient along the $x$ direction being constant. In addition, $v\sim 0$ and $w(y)$ is linear (Couette flow), $w=-Wy/h$. One concludes that, far from the edge of the blade ($x\ll -\ell$), $\left| \Sigma \right | \sim (\eta/h) \sqrt{3(4U^2+W^2)}$ and from Eq.~\ref{eqn : critere}, the thickness of the liquid layer along the blade is
\begin{equation}
h=\ell \sqrt{1+4\zeta^2}.
\end{equation}
Eqs.~\ref{eqn : incompressibility}-\ref{eqn : stokes w} and Eq.~\ref{eqn : critere} are solved using the finite-element method, implemented in the open-source finite-element library FEniCS~{\citep{Fenics2012}}. We adopt a set of units such that $\ell=1$ and $W=1$. Assuming reflectional symmetry, we consider domains $\Omega$ defined by
\begin{equation}
(x,y)\in \Omega \Leftrightarrow 0 \le y \le h \mbox{ and } x_{min} \le x \le \Gamma(y)
\label{eqn : Gamma}
\end{equation}
$\Gamma(y)$ is a decreasing function for $y\in[0,h]$. $x_{min}$ is the lower $x$ coordinates considered in the simulation ($x_{min}<\Gamma(h)$, see Fig.~\ref{fig : mesh}). Boundary conditions are detailed in Fig.~\ref{fig : mesh}.
\begin{figure}[!h]
\begin{center}
\includegraphics[width=0.4\textwidth]{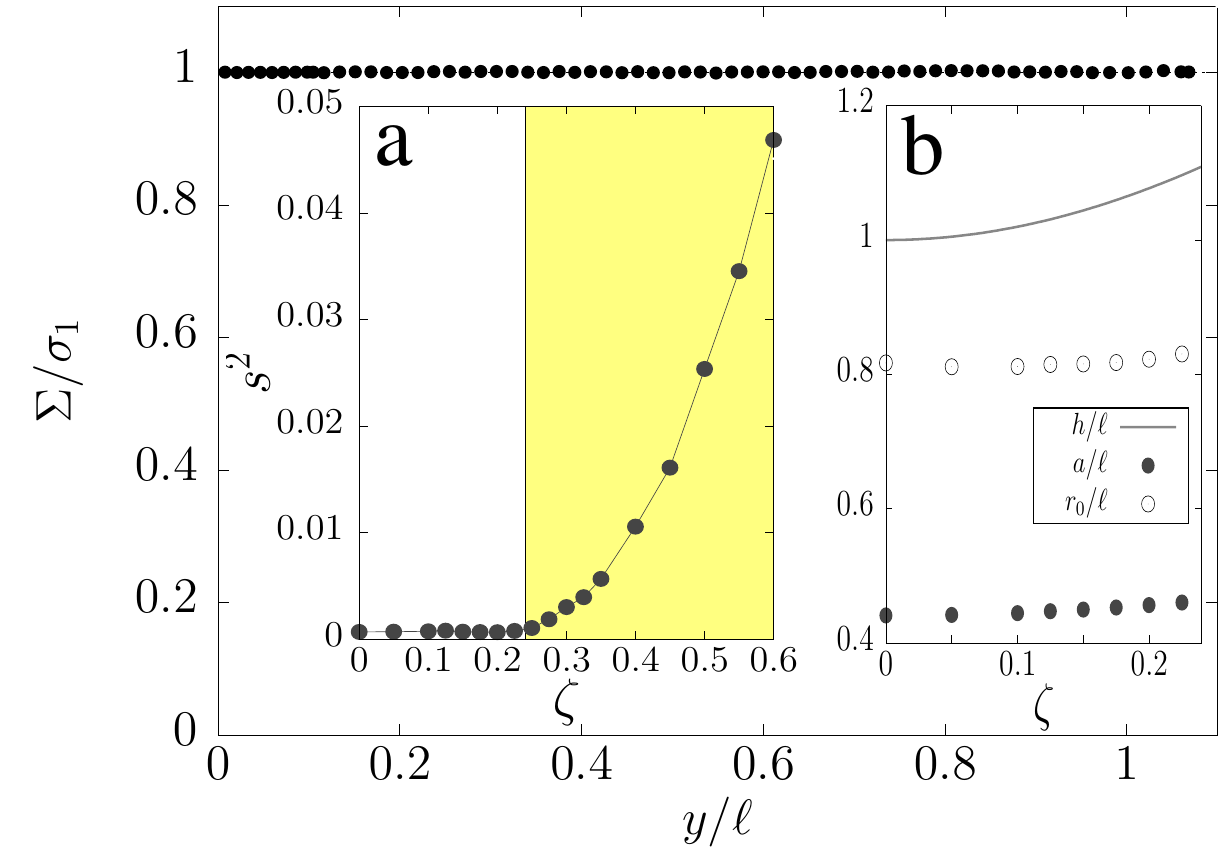}
\end{center}
\caption{Reduced stress $\Sigma/\sigma_1$ at the gel surface for $\zeta=0.2$, as a function of the reduced coordinate $y/\ell$, computed for the surface equation resulting from the fitting procedure aiming to approach boundary condition Eq.~\ref{eqn : critere}, with $(n,m)=(9,40)$ and $x_{min}=-5\ell$.
 (a) Variance $s^2$ calculated for the interface $\Gamma$ that best fits condition Eq.~\ref{eqn : critere} (with $(n,m)=(9,40)$ and $x_{min}=-5\ell$).
(b) $h$, $a$, and $r_0$ rescaled by $\ell$ as a function of $\zeta$.
} \label{fig : fit}
\end{figure}
We take for $\Gamma(y)$ a series expansion in the form
\begin{equation}
  \Gamma(y)=\sum_{k=0}^{n-1}\alpha_{2k}(y/h)^{2k}+\alpha_n\sum_{k=n}^m(y/h)^{2k}.
  \label{eqn : trial function}
\end{equation}
Coefficients $\alpha_{2k}$ are fitted in order to fulfill Eq.~\ref{eqn : critere}. The fits are done by carrying out a systematic exploration of the coefficients $\alpha_k$ and by minimizing the variance $s^2=\frac{1}{0.95h}\int_0^{0.95h} \frac{(\Sigma(y)-\sigma_1)^2}{\sigma^2_1}\mathrm{d}y$. The last sum in Eq.~\ref{eqn : trial function} has no significant effect on the best $\Gamma(y)$ except for $y\sim h$: neither a change in $n$ (provided that $n>8$) nor a change in $m$ (provided that $m>30$) has an effect on the values found for $\alpha_0$ or $\alpha_2$.   

Figure \ref{fig : fit} gives an example, for $\zeta=0.2$, of the reduced stress $\Sigma/\sigma_1$ calculated after the fitting procedure.  As required, $\Sigma/\sigma_1$ is almost constant and equal to one. Figure \ref{fig : fit} (a) shows that the variance $s^2$ calculated for the best fit starts to increase rapidly beyond a threshold value of $\zeta$,  $\zeta^*\simeq 0.24$. This indicates that solution of Eq.~\ref{eqn : critere} exists only for $\zeta<\zeta^*$, i.e. for $U< 0.24W$, suggesting that steady states exist only if the sliding velocity is large enough compared to the normal velocity, $W>U/0.24$.

The reduced normal distance $a/\ell=\alpha_0$ and the reduced radius of curvature $r_0/\ell=h^2/(2\alpha_2^2)$ are plotted as a function of $\zeta$ (for $\zeta<\zeta^*$) in  Fig.~\ref{fig : fit} (b). The rate of increase of $r_0/\ell$ as a function of $\zeta$ is lower than the rate for $h/\ell$: the profile of the notch is less blunt as $\zeta$ increases.
\begin{figure}[!h]
\begin{center}
\includegraphics[width=0.5\textwidth]{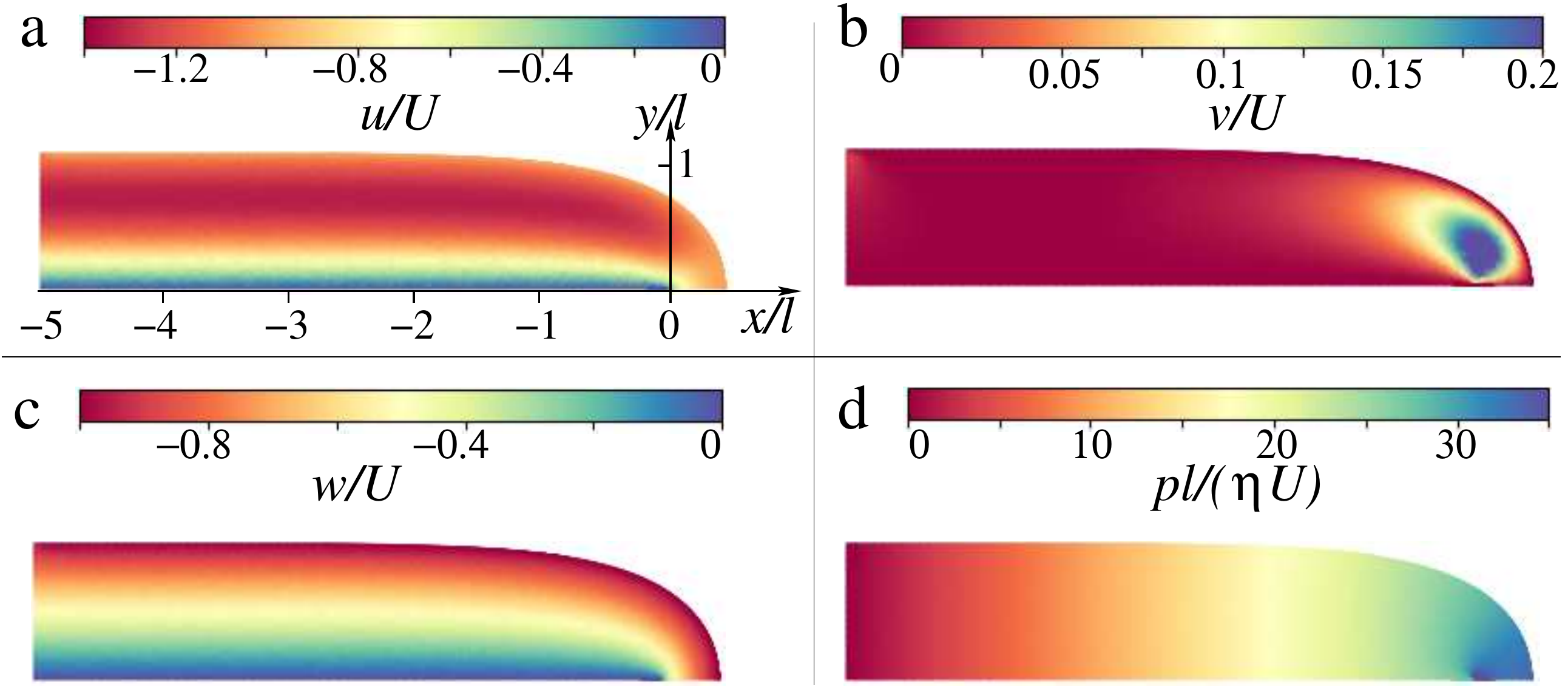}
\end{center}
\caption{Reduced components of the velocity $u$ (a), $v$ (b), and $w$ (c), and pressure $p$ (d), computed from Eqs.~\ref{eqn : incompressibility}-\ref{eqn : stokes w} and Eq. \ref{eqn : critere} using the finite element method, calculated for $\zeta=0.2$.} \label{fig : fem}
\end{figure}
Velocity components and pressure computed for $\zeta=0.2$ are shown in Fig.~\ref{fig : fem}.\\

To summarize, the cut made by pressing and sliding a sharp wedge on a soft material has been described by considering the viscous liquid layer surrounding the wedge. This layer results from the transformation of the soft solid to a liquid when the applied stress exceeds a prescribed value. The stress exerted on the wedge is transmitted to the soft solid thought the liquid layer, hence the prevailing role of the wedge velocities in the cutting process.

No steady state in the melting of the solid can be induced by a pure normal indentation of the sharp wedge ($W=0$). Indeed, a steady regime requires a large enough ratio of the sliding velocity to the normal velocity. The maximal value of the normal velocity (the cutting speed) is proportional to the tangential velocity, $U_{max}\simeq 0.24W$. Hence, the maximum cutting speed is directly related to the sliding velocity: quicker dicing requires higher tangential velocity. 

For given imposed normal and tangential velocities in a steady regime, the condition that the critical stress of the gel has to be reached fixes the shape of the transition zone. The thickness of the fluid layer far behind the edge of the blade, the thickness in the direction normal to the edge, and also the minimal radius of curvature of the transition zone, have been computed. These lengths are determined by the fluid-gel interaction, and not by the elasticity of the gel alone.

Whether the deformations of the solid phase are large or small does not matter in the theory, the important property  being that the gel remains elastic below a critical stress is reached. In that sense, the theory fundamentally departs from standard theories of fracture mechanics that are based on the calculation of elastic deformations together with an estimation of the energy release rate taking place in the plastic zone.

Extending the theory introduced here to unsteady states would be useful to capture the nucleation stages, to unfold the cases steady states do not exist (e.g., normal dicing), and to explain how a tangential vibration can improve dicing, as evidenced in the puncture of soft gels \cite{chakrabarti2016,Kundan2019} or practiced in surgery \cite{Giovannini2018,Wang2019}.

Experiments have now to be carried out in order to test these predictions, for instance by characterizing geometrical properties of the liquid layer and by evidencing the minimal tangential to normal velocity ratio for a steady regime in the cutting.


\end{document}